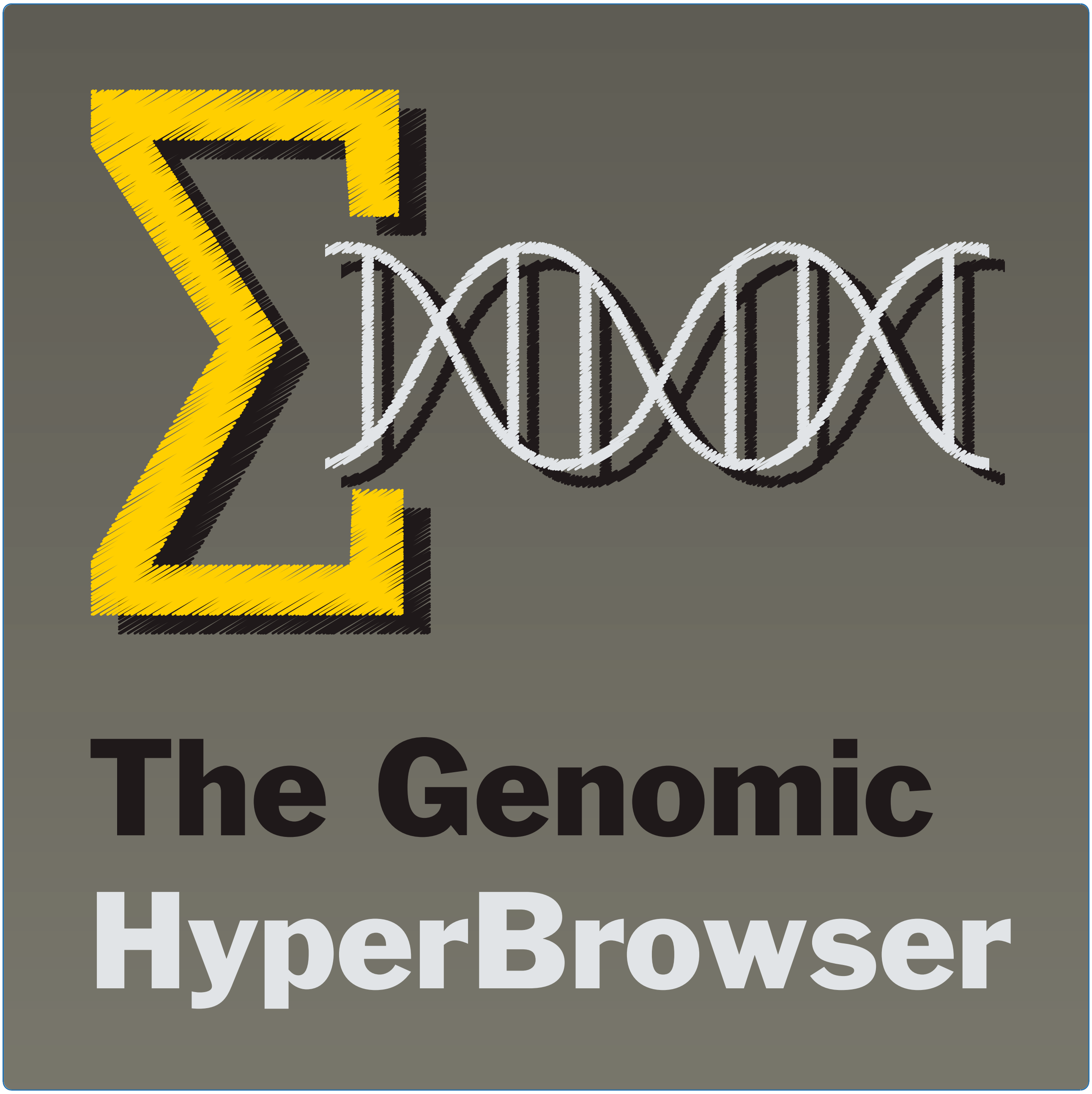

# The Genomic HyperBrowser: inferential genomics at the sequence level

Sandve *et al.*

# The Genomic HyperBrowser: inferential genomics at the sequence level

Geir K Sandve[1], Sveinung Gundersen[2], Halfdan Rydbeck[1,3,5], Ingrid K Glad[4], Lars Holden[3], Marit Holden[3], Knut Liestøl[1,5], Trevor Clancy[2], Egil Ferkingstad[3], Morten Johansen[6], Vegard Nygaard[6], Eivind Tøstesen[6], Arnoldo Frigessi[3,7], Eivind Hovig[1,2,3,6*]

## Abstract
The immense increase in the generation of genomic scale data poses an unmet analytical challenge, due to a lack of established methodology with the required flexibility and power. We propose a first principled approach to statistical analysis of sequence-level genomic information. We provide a growing collection of generic biological investigations that query pairwise relations between tracks, represented as mathematical objects, along the genome. The Genomic HyperBrowser implements the approach and is available at http://hyperbrowser.uio.no.

## Rationale
The combination of high-throughput molecular techniques and deep DNA sequencing is now generating detailed genome-wide information at an unprecedented scale. As complete human genomic information at the detail of the ENCODE project [1] is being made available for the full genome, it is becoming possible to query relations between many organizational and informational elements embedded in the DNA code. These elements can often best be understood as acting in concert in a complex genomic setting, and research into functional information typically involves integrational aspects. The knowledge that may be derived from such analyses is, however, presently only harvested to a small degree. As is typical in the early phase of a new field, research is performed using a multitude of techniques and assumptions, without adhering to any established principled approaches. This makes it more difficult to compare, reproduce and realize the full implications of the various findings.

The available toolbox for generic genome scale annotation comparison is presently relatively small. Among the more prominent tools are those embedded within the genome browsers, or associated with them, such as Galaxy [2], BioMart [3], EpiGRAPH [4] and UCSC Cancer Genomics Browser [5]. BioMart at this point mostly offers flexible export of user-defined tracks and regions. Galaxy provides a richer, text-centric suite of operations. EpiGraph presents a solid set of statistical routines focused on analysis of user-defined case-control regions. The recently introduced UCSC Cancer Genomics Browser visualizes clinical omics data, as well as providing patient-centric statistical analyses.

We have developed novel statistical methodology and a robust software system for comparative analysis of sequence-level genomic data, enabling integrative systems biology, at the intersection of genomics, computational science and statistics. We focus on inferential investigations, where two genomic annotations, or tracks, are compared in order to find significant deviation from null-model behavior. Tracks may be defined by the researcher or extracted from the sizable library provided with the system. The system is open-ended, facilitating extensions by the user community.

## Results
### Overview
Our system is based on an abstract representation of generic genomic elements as mathematical objects. Hypotheses of interest are translated into mathematical relations. Concepts of randomization and track structure preservation are used to build complex problem-specific null models of the relation between two tracks. Formal inference is performed at a global or local scale, taking confounder tracks into account when necessary (Figure 1).

* Correspondence: ehovig@ifi.uio.no
[1]Department of Informatics, University of Oslo, Blindern, 0316 Oslo, Norway
Full list of author information is available at the end of the article

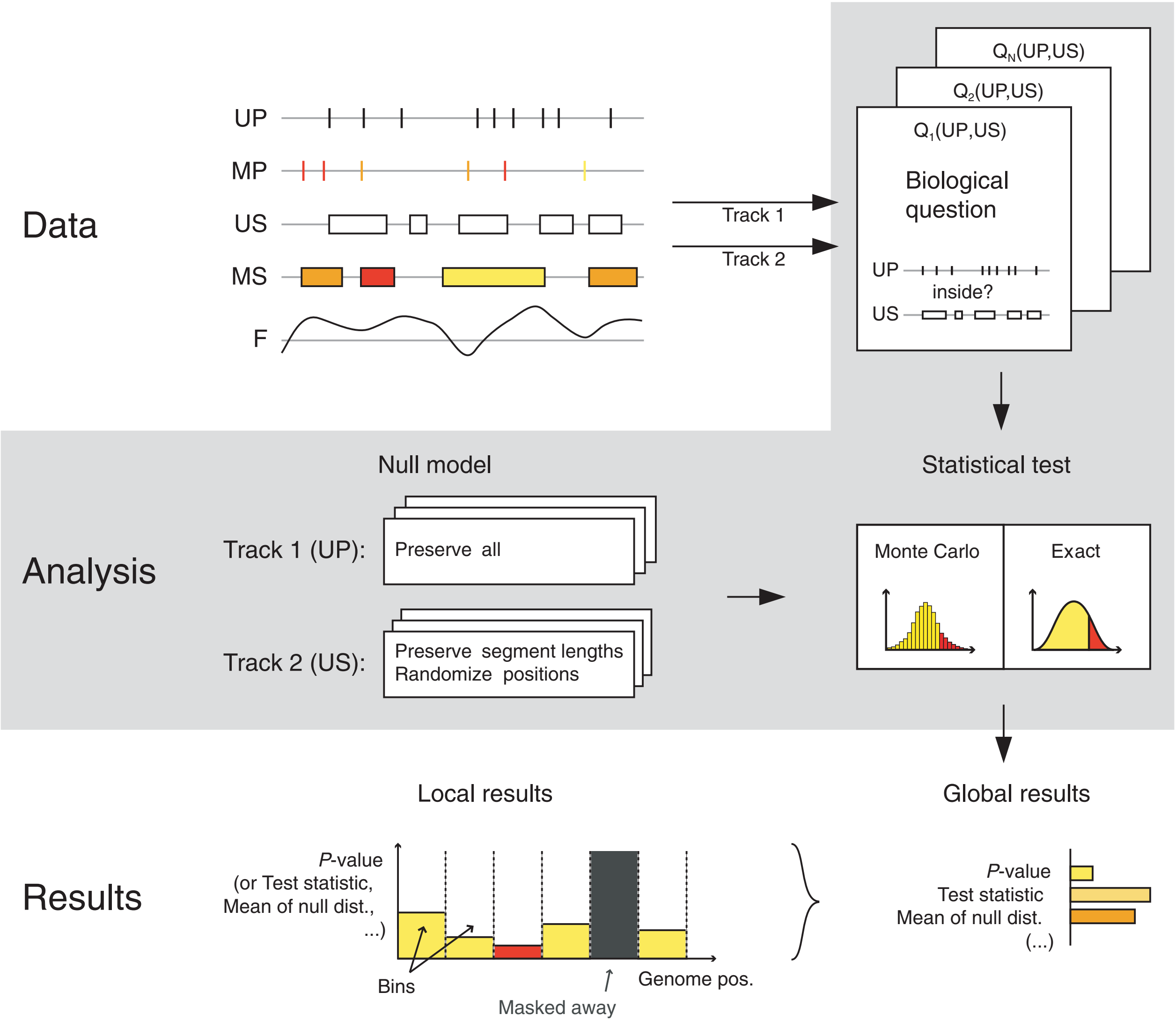

**Figure 1 Flow diagram of the mathematics of genomic tracks**. Genomic tracks are represented as geometric objects on the line defined by the base pairs of the genome sequence: (unmarked (UP) or marked (MP)) points, (unmarked (US) or marked (MS)) segments, and functions (F). The biologist identifies the two tracks to be compared, and the Genomic HyperBrowser detects their type. The biological question of interest is stated in terms of mathematical relations between the types of the two tracks. The relevant questions are proposed by the system. The biologist then selects the question and needs to specify the null hypothesis. For this purpose she is called to decide about what structures are preserved in each track, and how to randomize the rest. Thereafter, the Genomic HyperBrowser identifies the relevant test statistics, and computes actual *P*-values, either exactly or by Monte Carlo testing. Results are then reported, both for a global analysis, answering the question on the whole genome (or area of study), and for a local analysis. Here, the area is divided into bins, and the answer is given per bin. *P*-values, test-statistic, and effect sizes are reported, as tables and graphics. Significance is reported when found, after correction for multiple testing.

### Abstract representation of genomic elements

A genome annotation track is a collection of objects of a specific genomic feature, such as genes, with base-pair-specific locations from the start of chromosome 1 to the end of chromosome Y. Tracks vary in biological content, but also in the form of the information they contain. A track representing genes contains positional information that can be reduced to 'segments' (intervals of base pairs) along the genome. A track of SNPs can be reduced to points (single base pairs) on the genome. The expression values of a gene, or the alleles of a SNP, are non-positional information parts and are attributed as 'marks' (numerical or categorical) to the corresponding positional objects, that is, segments or points. Finally, a track of DNA melting assigns a temperature to each base pair, describing a 'function' on the genome. We thus define five genomic types: unmarked points (UP), marked points (MP), unmarked segments (US), marked segments (MS) and functions (F). These five types completely represent every one-dimensional geometry with marks.

### Catalogue of investigations

We translate biological hypotheses of interest into a study of mathematical relations between genomic tracks, leading to a large collection of possible generic investigations.

Consider the relation between histone modifications and gene expression, as investigated by visual inspection in [6] (Figure S1 in Additional file 1). The question is whether the number of nucleosomes with a given histone modification (represented as type UP), counted in a region around the transcription start site (TSS) of a gene, correlates with the expression of the gene. The second track is represented as marked segments (MS). This study of histone modifications and gene expressions can then be phrased as a generic investigation between a pair of tracks (T1, T2) of type UP and MS: are the number of T1 points inside T2 segments correlated with T2 marks? Figure 2 shows the results when repeating this analysis for all histone modifications studied in [6], and different regions around the TSS. See Section 1 in Additional file 1 for a more detailed example investigation, analyzing the genome coverage by different gene definitions.

In the context of the catalogue of investigations, the genomic types are minimal models of information content. In the above example, nucleosome modifications are only used for counting, and thus considered unmarked points (UP), even though they are typically represented in the file system as marked or unmarked segments. As the gene-related properties of interest are the genome segments in which the nucleosomes are counted, as well as the corresponding gene expression values (marks), T2 is of the type marked segments (MS). The choice of genomic type clarifies the content of a track, and also restricts which analyses are appropriate. Investigations regarding the length of the elements of a track are, for instance, relevant for genes, but not for SNPs and DNA melting temperatures.

The five genomic types lead to 15 unordered pairs (T1, T2) of track type combinations, with each combination defining a specific set of relevant analyses. For instance, the UP-US combination defines several investigations of potential interest: are the T1 points falling inside the T2 segments more than expected by chance? Do the points accumulate more at the borders of the segments, instead of being spread evenly within? Do the points fall closer to the segments than expected? A growing collection of abstract mathematical versions of biological questions is provided. We have currently implemented 13 different analyses, filling 8 of the 15 possible combinations of track types (see Additional file 2 for mathematical details). Note that information reduction of a track to a simpler type (for example, segments to points) may open up additional analytical opportunities, and are handled dynamically by the system - for example, by treating segments as their middle points.

### Global and local inference

A global analysis investigates if a certain relation between two tracks is found in a domain as a whole. A local analysis is based on partitioning the domain into smaller units, called bins, and performing the analysis in each unit separately. Local analysis can be used to investigate if and where two tracks display significant concordant or discordant behavior, and thus be used to generate hypotheses on the existence of biological mechanisms explaining such perturbations. Local investigations may also be used to examine global results in more detail. The length of each bin defines the scale of the analysis. Inference is then based on the computation of *P*-values, locally in each bin, or globally, under the null model.

To illustrate the value of local analysis, we consider viral integration events in the human genome. These may result in disease and may also be a consequence of retroviral gene therapy. Derse *et al.* [7] examined integration for six types of retroviruses, with different viral integrases, thus having different integration sites (type UP). Using these data, we asked whether there are hotspots of integration inside 2-kb flanking regions of predicted promoters (type US), that is, whether and where the points are falling inside the segments more than expected by chance. Figure 3 displays the hotspots as calculated *P*-values in bins across the genome, using the subset of murine leukemia virus (MLV) sites. We find locations of increased integration, thus generating hypotheses on the role of integration site sequences and their context.

Local analysis may be used to avoid drawing incorrect conclusions from global investigations. Consider the repressive histone modification H3K27me3 as studied in [8]. Data from ChIP-chip experiments on mouse chromosome 17 were analyzed, finding that H3K27me3 falls in domains that are enriched in short interspersed nuclear element (SINE) and depleted in long interspersed nuclear element (LINE) repeats. Using the line of enquiry raised in [8], we asked whether H3K27me3 regions (type US) significantly overlap with SINE repeats (type US), but here using formal statistical testing at the base pair level. The chosen null model only allows local rearrangements of genomic elements (for more detail, see next section). This preserves local biological structure, but allows for some controlled level of randomness.

Performing this test globally on the whole chromosome 17 leads to rejection of the null hypothesis ($P = 10^{-4}$),

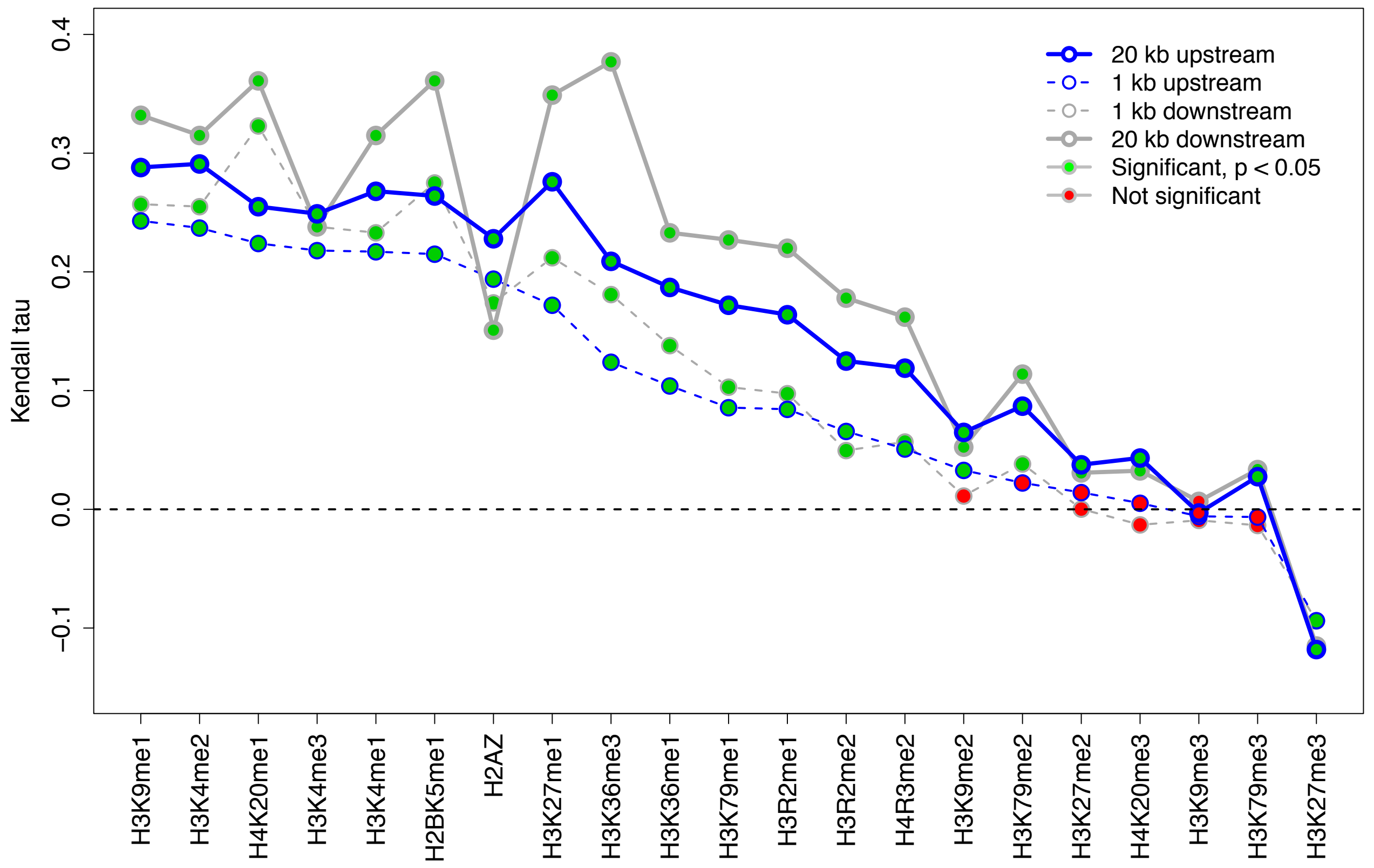

**Figure 2 Gene regulation by histone modifications**. The correlation between occupancy of 21 different histone modifications and gene expression within 4 different regions around the TSS (up- and downstream, 1 and 20 kb), sorted by correlation in 1-kb upstream regions. Sixteen of 21 histone modifications show significant correlation in 1-kb upstream regions, while inspection of the actual value of Kendall's tau (Table S1 in Additional file 1) shows very little effect size for 6 of these 16 (<0.1).

in line with [8]. However, a local analysis leads to a deeper understanding. At a 5-Mbp scale, no significant findings were obtained in any of the 19 bins (10% false discovery rate (FDR)-corrected). The frequency of H3K27me3 segments varies considerably along chromosome 17 (Figure S2 in Additional file 1), which may cause the observed discrepancy between local and global results.

## Precise specification of null models

A crucial aspect of an investigation is the precise formalization of the null model, which should reflect the combination of stochastic and selective events that constitutes the evolution behind the observed genomic feature.

Consider again the example of H3K27me3 versus repeating elements. In the chosen null model, we preserved the repeat segments exactly, but permuted the positions of the H3K27me3 segments, while preserving segment and intersegment lengths. We then computed the total overlap between the segments, and used a Monte Carlo test to quantify the departure from the null model. The effect of using alternative null models is shown in Table 1. The null model examined in the first column, which does not preserve the dependency between neighboring base pairs, produces lower *P*-values. Unrealistically simple null models may thus lead to false positives. In fact, two simulated independent tracks may appear to have a significant association if their individual characteristics are not appropriately modeled (Section 2 in Additional file 1). In this example, the choice between the biologically more reasonable null models is difficult. The two other columns of Table 1 include models that preserve more of the biological structure. The fact that these models do not lead to clear rejection of the null hypotheses suggests that we in this case lack strong evidence against the null hypothesis. Thus, examining the results obtained for a set of different null models may often contribute important information. The null model should reflect biological realism, but also allow sufficient variation to permit the construction of tests. A set of simulated synthetic tracks is provided as an aid for assessing appropriate null models (Additional file 3).

The Genomic HyperBrowser allows the user to define an appropriate null model by specifying (a) a preservation

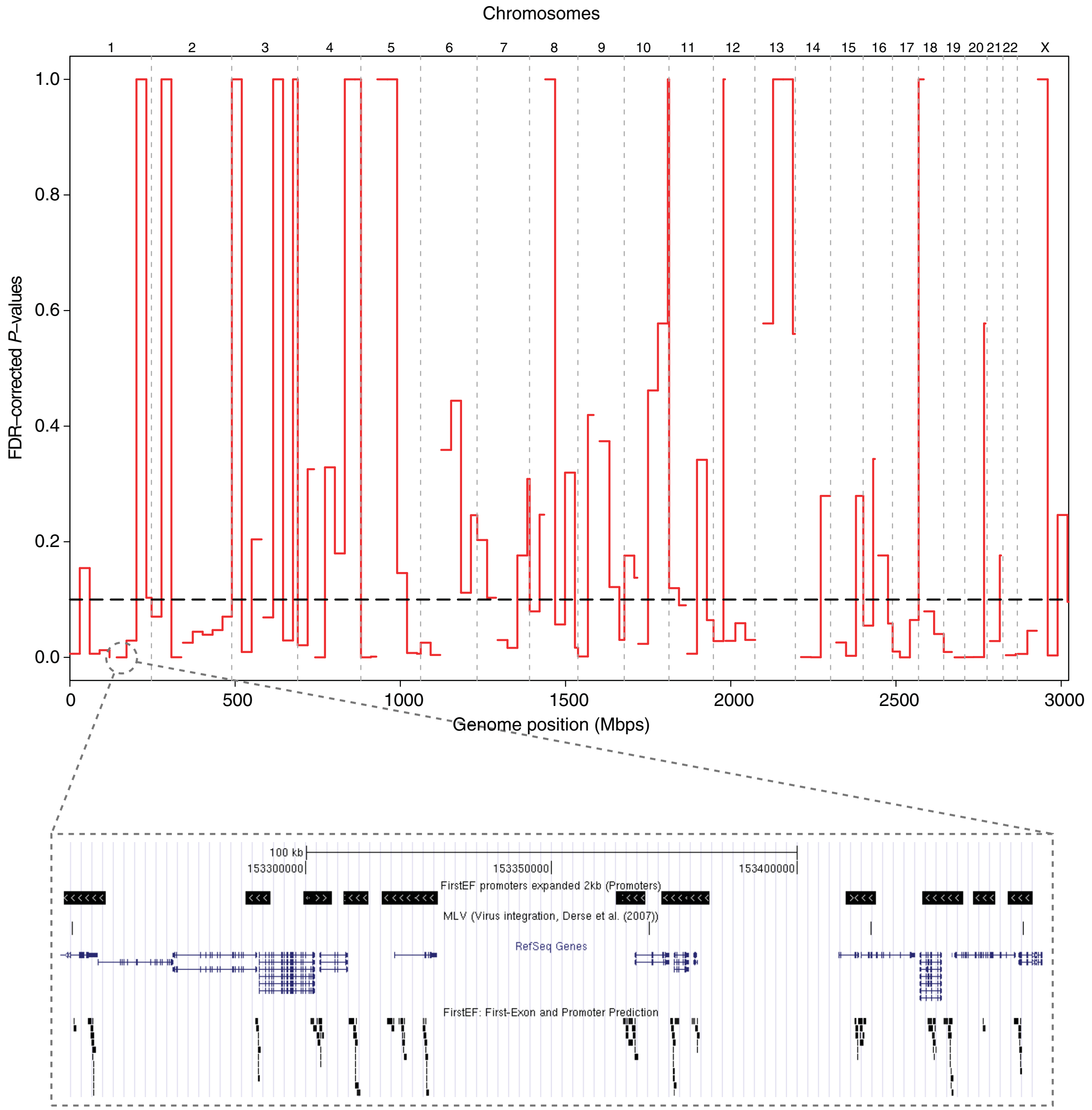

**Figure 3 Viral integration sites**. Plot of false discovery rate (FDR)-adjusted *P*-values along the genome, in 30-Mbp bins. Small *P*-values indicate regions where murine leukemia virus (MLV) integrates inside 2-kb regions around FirstEF promoters more frequently than by chance. The FDR cutoff at 10% is shown as a dashed line. The inset of a local area (chromosome 1:153,250,001-153,450,000) indicates FirstEF promoters expanded by 2 kb in both directions, MLV integration sites, RefSeq genes, and unflanked FirstEF sites.

rule for each track, and (b) a stochastic process, describing how the non-preserved elements should be randomized. Preservation fixes elements or characteristics of a track as present in the data. For each genomic type, we have developed a hierarchy of less and less strict preservation rules, starting from preserving the entire track exactly (Section 3 in Additional file 1). For example, these preservation options for unmarked segments can be assumed: (i) preserve all, as in data; (ii) preserve segments and intervals between segments, in number and length, but not their ordering; (iii) preserve only the segments, in number and length, but not their position; (iv) preserve only the number of base pairs in segments, not segment position or number. Depending

**Table 1 Significant bins of the overlap test between H3K27me3 segments and SINE repeats under various null models**

| Tracks to randomize | Preserve total number of base pairs covered | Preserve segment lengths, but randomize position | Preserve segment and intersegment lengths, but randomize positions |
|---|---|---|---|
| H3K27me3 | 10/19 | 1/19 | 0/19 |
| SINE | 10/19 | 5/19 | 4/19 |
| H3K27me3 and SINE | 10/19 | 5/19 | 4/19 |

The number of significant bins of the overlap test between H3K27me3 segments and SINE repeats under different preservation and randomization rules for the null model. The test was performed in 19 bins on mouse chromosome 17, with the MEFB1 cell line. (Use of the MEFF cell line gave similar results; Table S2 in Additional file 1). In this case, less preservation of biological structure leads to smaller *P*-values. Also, randomizing the SINE track gave smaller *P*-values than randomizing the H3K27me3 track (or both).

on the test statistic T, the level of preservation and the chosen randomization, *P*-values are computed exactly, asymptotically or by standard or sequential Monte Carlo [9,10].

### Confounder tracks

The relation between two tracks of interest may often be modulated by a third track. Such a third track may act as a confounder, leading, if ignored, to dubious conclusions on the relation between the two tracks of interest.

Consider the relation of coding regions to the melting stability of the DNA double helix. Melting forks have been found to coincide with exon boundaries [11-15]. Although few studies have reported statistical measures of such correlation [11], the correlation is confirmed by a straightforward investigation. Tracks (type F) representing the probabilities of melting fork locations [16] in *Saccharomyces cerevisiae*, were compared to tracks containing all exon boundaries (Figure 4). We asked if the melting fork probabilities (P) were higher than expected at the exon boundaries (E) than elsewhere. In the null model, the function was conserved, while points were uniformly randomized in each chromosome. Monte Carlo testing was carried out on the chromosomes separately, giving *P*-values <0.0005 (Table S3 in Additional file 1). In the absence of a confounder, it is thus tempting to conclude that there is an interesting relation between DNA melting and coding regions, for which functional implications have been previously discussed [15,17,18].

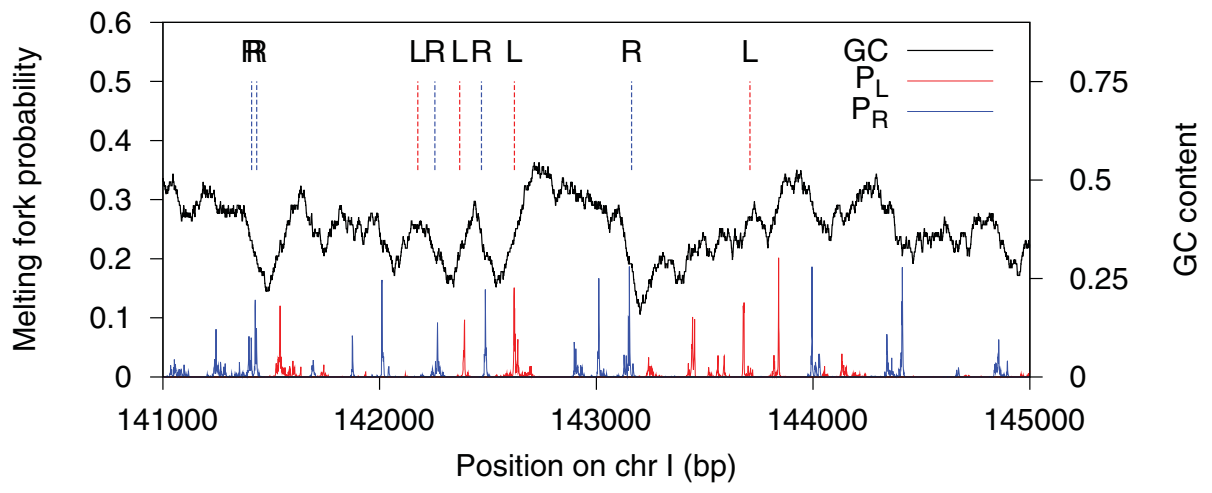

**Figure 4 Comparison of exon boundary locations and melting fork probability peaks**. Independent analyses were carried out on left and right exon boundaries as compared to left- and right-facing melting forks, respectively. In the upper part, dashed vertical lines indicate left (L, red) and right (R, blue) exon boundaries. In the lower part, probabilities of left- and right-facing melting forks appear as red and blue peaks, respectively. The black curve shows the GC content in a 100-bp sliding window (values on right axis).

An alternative view is that the GC content, being higher inside exons than outside, contains information about exon location that is simply carried over, or decoded, by a melting analysis, thus acting as a confounder. We have developed a methodology to investigate such situations further. Non-preserved elements of a null model can be randomized according to a non-homogeneous Poisson process with a base-pair-varying intensity, which can depend on a third (or several) modulating genomic tracks [19,20]. We have defined an algebra for the construction of intensities, where tracks are combined, to allow rich and flexible constructions of randomness (see Materials and methods).

To investigate the influence of GC content on the exon-melting relation, we first generated a pair of custom tracks (type F), assigning to each base the value given by the GC content in the 100-bp left and right flanking regions, respectively, weighted by a linearly decreasing function. These two functions were used, together with the exon boundary track, to create an intensity curve proportional to the probability of exon points, given GC content (see Materials and methods). When performing the same analysis as before, but now using the null model based on this intensity curve (rather than assuming uniformity), a significant relationship was found in only one yeast chromosome (Table S3 in Additional file 1). In conclusion, there is a melting-exon relationship in yeast, but it may simply be a consequence of differences in GC content at the exon boundaries (high GC inside, low GC outside), which may exist for biological reasons not involving melting fork locations.

### Resolving complexity: system architecture

The Genomic HyperBrowser is an integrated, open-source system for genome analysis. It is continually evolving, supporting 28 different analyses for significance testing, as well as 62 different descriptive statistics. The system currently hosts 184,500 tracks. Most of these represent literature-based information, previously mostly utilized in network-based approaches [21]. As natural language based text mining allows for the identification of a wide variety of biological entities, we have generated tracks representing genomic locations associated with terms for the complete gene ontology tree, all Medical Subject Heading (MeSH) terms, chemicals, and anatomy.

The system is implemented in Python [22], a high-level programming language that allows fast and robust software development. A main weakness of Python compared to languages like C++ is its slower performance. Thus, a two-level architecture has been designed. At the highest level, Python objects and logic have been used extensively to provide the required flexibility. At the base-pair level, data are handled as low-level vectors, combining near-optimal storage with efficient indexing, allowing the use of vector operations to ensure speed. Interoperability with standard file formats in the field [23] is provided by parallel storage of original file formats and preprocessed vector representations. To reduce the memory footprint of analyses on genome-wide data, an iterative divide-and-conquer algorithm is automatically carried out when applicable. A further speedup is achieved by memoizing intermediate results to disk, automatically retrieving them when needed for the same or different analyses on the same track(s) at any subsequent time, by any user.

The system provides a web-based user interface with a low entry point. However, the complex interdependencies between the large body of available tracks, a number of syntactically different analyses, and a range of choices for constructing null models, all pose challenges to the concepts of simplicity and ease of use. In order to simplify the task of making choices, a step-wise approach has been implemented, displaying only the relevant options at each stage. This guided approach hides unnecessary complexities from the researcher, while confronting her with important design choices as needed. We rely on a dynamic system to infer appropriate options, aiding maintenance. The list of selectable tracks is based on scans of available files on disk. The list of relevant questions is based on short runs of all implemented analyses, using a minimal part of the actual data from the selected tracks. For each analysis, a set of relevant options is defined. The dynamics of the system also provides automatic removal of analyses that fail to run, enhancing system robustness.

Allowing extensibility along with efficiency and system dynamics is a challenge. The complexities of the software solutions are hidden in the backbone of the system, simplifying coding of statistical modules. Each module declares the data types it supports and which results are needed from other modules. The backbone automatically checks whether the selected tracks meet the requirements, and if so, makes sure the intermediate computations are carried out in correct order. Redundant computations are avoided through the use of a RAM-based memoization scheme. The system also provides a component-based framework for Monte Carlo tests, where any test statistic can be combined with any relevant randomization algorithm, simplifying development. In addition, a framework for writing unit and integration tests [24] is included. Further details on the system architecture are provided in Section 4 in Additional file 1.

### Step-by-step guide to HyperBrowser analysis

One of the main goals of the Genomic HyperBrowser is to facilitate sophisticated statistical analyses. A range of textual guides and screencasts are available in the help section at the web page, demonstrating execution of various analyses, how to work with private data, and more. To give an impression of the user experience, we here provide a step-by-step guide to the analysis of broad local enrichment (BLOC) segments versus SINE repeats, as discussed in the section on 'Precise specification of null models'.

First, we open 'hyperbrowser.uio.no' in a web browser and we select the 'Perform analysis' tool under 'The Genomic HyperBrowser' in the left-hand menu. We select the mouse genome (mm8) and continue to select tracks of interest. As the first track, we select 'Chromatin'-'Histone modifications'-'BLOC segments'-'MEFB1'. These are the BLOC segments according to the algorithm of Pauler *et al.* [8] for the MEFB1 cell line. As the second track, we select 'Sequence'-'Repeating elements'-'SINE'. Now that both tracks have been selected, a list of relevant investigations is presented in the interface (that is, investigations that are compatible with the genomic types of the two tracks: US versus US). We select the question of 'Overlap?' in the 'Hypothesis testing' category, and the options relevant for this analysis are subsequently displayed in the interface. The different choices for 'Null model' will produce the various numbers in Table 1 (six different choices are directly available from the list. The other variants can be achieved by reversing the selection order of the tracks). The original BLOC paper [8] focused on chromosome 17. We want to perform a local analysis along this chromosome, avoiding the first three megabases that are centromeric. Under 'Region and scale' we thus choose to 'Compare in' a custom specified region, writing 'chr17:3m-' as 'Region of the genome' and writing '5 m' (5 megabases) as 'Bin size'. Clicking the 'Start analysis' button will then perform an appropriate statistical test according to the selected null model assumption, and output textual and graphical results to a new Galaxy history element. Figure 5a shows the user interface covering all selections above and Figure 5b shows the answer page that results from this analysis.

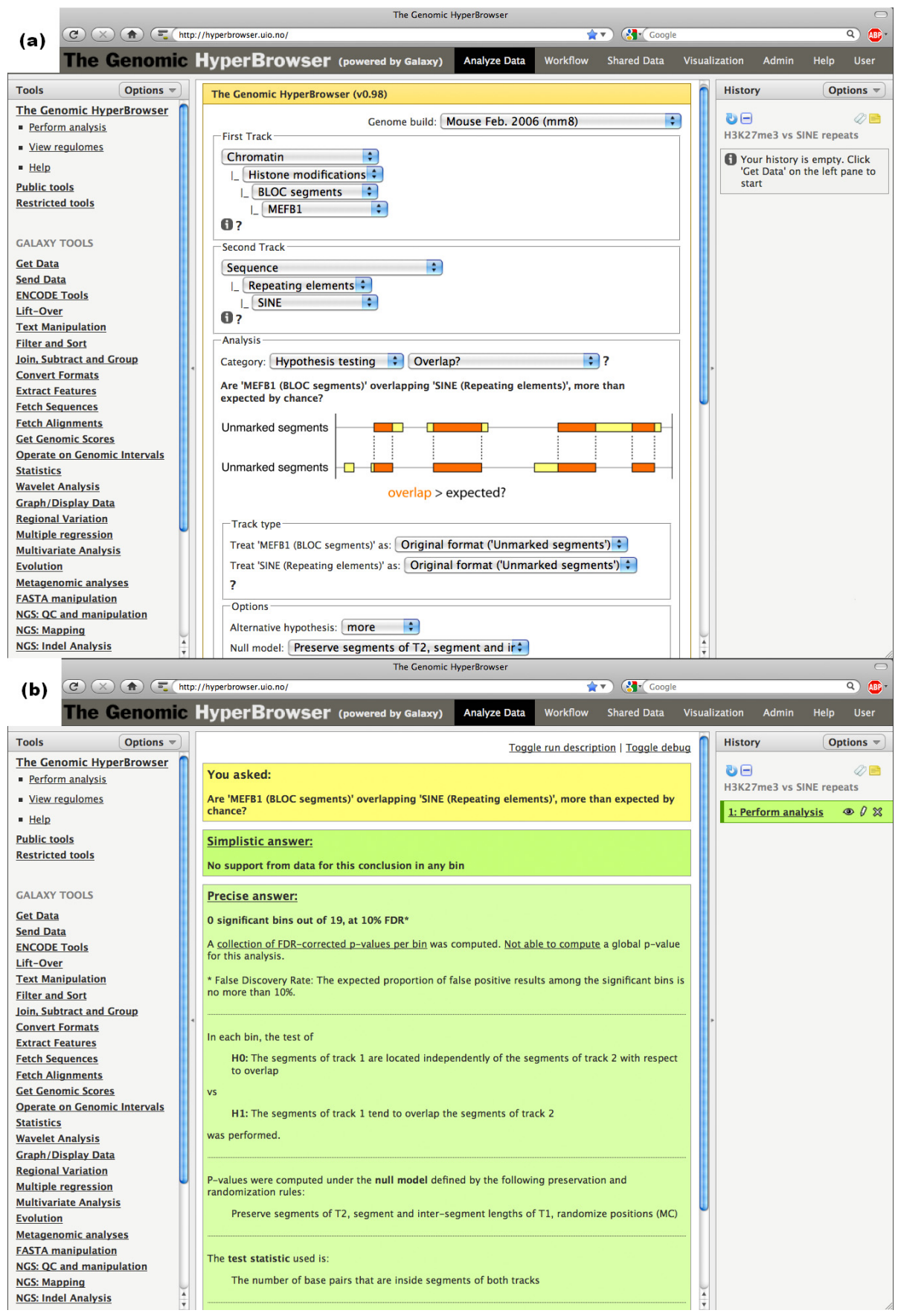

**Figure 5 Screenshots of the Genomic HyperBrowser. (a)** Screenshot of the main interface for selecting analysis options. The selections for the example relating H3K27me3 BLOCs to SINE repeats have been pre-selected. In the interface, the user selects a genome build followed by two tracks. A list of relevant investigations is then presented, based on the genomic types of the two tracks. After selecting an investigation, the interface presents the user with a choice of null models, alternative hypotheses and other relevant options. **(b)** Screenshot of the results of the analysis. The question asked by the user is presented at the top, in this case: 'Are 'MEFB1 (BLOC segments)' overlapping 'SINE (Repeating elements)' more than expected by chance?' A first, simplistic answer is then presented: 'No support from data for this conclusion in any bin'. A more precise answer follows, detailing any global *P*-values, a summary of local FDR-corrected *P*-values, the particular set of null and alternative hypotheses tested, in addition to a legend of the test statistic that has been used. Further links to a PDF file containing the statistical details of the test, and to more detailed tables of relevant statistics for both the global and the local analysis are also included. The global result table also includes links to plots and export opportunities for the individual statistics.

This example assumed the BLOC segments were already in the system. If not, they could simply be uploaded to the Galaxy history and then selected in the first track menu as '– From history (bed, wig) –'-'[your BLOC history element]'. For information on how to use the Galaxy system, we refer to the Galaxy web site [25].

## Discussion

The current leap in high-throughput sequencing technology is opening the way for a range of genome-wide annotations beyond the presently abundant gene-centric data. Not least, chromatin-related data are becoming increasingly important for understanding higher-level organization and regulation of the genome [26].

As is typical for a subfield that has not reached maturation, analysis of new massive sequence-level data is performed on a per-project basis. For instance, a paper on the ENCODE project describes how inference can be done by Monte Carlo testing, sampling bins for one of the real tracks at random genome locations under the null hypothesis [1]. Independently, a newer study of histone modifications instead permuted bins of data for one of the tracks [27]. Although genomic visualization tools have been available for several years, few generic tools exist for inference at the sequence level.

The following aspects distinguish our work from currently available systems. First, we focus on genomic information of a sequential nature, that is, with specific base-pair locations on a genome, and thus not restricted to only genes. Second, it focuses on the comparison of pairs of genomic tracks, possibly taking others into account through the concept of intensity tracks. Third, all comparisons are performed using formal statistical testing. Fourth, we provide analyses on any scale, from genome-wide studies to miniature investigations on particular loci. Fifth, we offer flexible choices of null models for exploration and choice where relevant. Finally, we provide a user interface where the user describes the data and the null models, while the system based on this chooses the appropriate statistical test. Comparing this to the EpiGRAPH and Galaxy frameworks, which we believe are the closest existing systems, we find that both require substantial technical expertise when choosing the correct analysis and options. EpiGRAPH is focused on a specific type of scenario that, according to our cataloguing, amounts to the comparison of unmarked points or segments versus categorically marked segments (with mark being case or control). Galaxy provides a simple user interface, is rich in tools for manipulating and analyzing datasets of diverse formats, but has little support for formal statistical testing. Note also that our system is tightly connected to Galaxy and can make use of all the tools provided within Galaxy.

We provide tools for abstraction and cataloguing of what we believe are typical questions of broad interest.

The abstractions of genomic data, the proposing of prototype investigations, and the careful attention given to null models simplifies statistical inference for a range of possible research topics. Our approach invites researchers to build relevant null models in a controlled manner, so that specific biological assumptions can be realistically represented by preservation, randomness and intensity based confounders. In addition, time used for repetitive tasks like file parsing and calculation of descriptive statistics may be significantly reduced.

Our system is highly extensible. The software is open source, inviting the community to add new investigations and tools. Attention has been given to component-based coding and simple interfaces, facilitating extensions of the system.

The highly specialized nature of many research investigations poses a major challenge for a generic system such as the one presented here. Even though a range of analyses and options are provided, chances are that at a given level of complexity, functionality beyond what is provided by a generic system will be needed. Still, the time and effort used to reach such a point may be shortened considerably, and it should in many cases be possible to meet demands through custom extensions.

Genomic mechanisms commonly involve more than two tracks, and the current focus on pair-wise interrogations is limiting. Our methodology allows the incorporation of additional tracks through the concept of an intensity track that modulates the null hypothesis, acting as a confounder. However, the investigation of genuine multi-track interactions is not yet possible within the system, as complex modeling and testing of multiple dependencies will be required.

Attention should be given to the trade-off between fine resolution and lack of precision. When large bins are considered, there may be too little homogeneity, while small bins may contain too little data. There is also an unresolved trade-off relating to preservation of tracks in null-hypotheses construction: too little preservation may give unrealistically small *P*-values, while too strong preservation may give too limited randomness.

On a more specific note, a set of tissue-specific analytical options would be beneficial with respect to many types of experimental data - for example, chromatin, expression and also gene subset tracks. Such options are now under development.

Novel sequencing technologies are instrumental in realizing the personalized genomes [28], and with them the task of identifying phenotype-associated information contained in each genome. An imminent challenge in understanding cellular organization is that of the three dimensions of the genome. While a number of genomes have been sequenced, and a number of important cellular elements have been mapped on a linear scale, the mapping of the three-dimensional organization of the DNA and chromatin in the nucleus is still only in its beginnings. Consequently, the impact of this organization on cell regulation is still largely unresolved. However, the advent of methods like Hi-C [29] permits detailed maps of three-dimensional DNA interactions to be combined with coarser methods of mapping of other elements. It appears that looking simultaneously at multiple scales seems important for understanding the dynamics of different functional aspects, from chromosomal domains down to the nucleosome scale. The need for taking multiple scales into account has recently been emphasized in both theoretical and analytical settings [30,31]. Consequently, statistical genomics needs to consider several scales when proper analytical routines are developed. Our approach is open to three-dimensional extensions, where the bins, which are flexibly selected in the system, will become three-dimensional volumes, and local comparison will be within each volume. What appears much more complex is the level of dependence of such volumes. But as the three-dimensional organization of the genome will become increasingly known, appropriate volume topologies will be possible, so that neighboring volumes representing three-dimensional contiguity may be used as a basis for statistical tests.

## Conclusions

By introducing a generic methodology to genome analysis, we find that a range of genomic data sets can be represented by the same mathematical objects, and that a small set of such objects suffice to describe the bulk of current data sets. Similarly, a range of biological investigations can be reduced to similar statistical analyses. The need for precise control of assumptions and other parameters can furthermore be met by generic concepts such as preservation and randomization, local analysis (binning) and confounder tracks.

Applying these ideas on a sample set of genomic investigations underlines that the generic concepts fit naturally to concrete analyses, and that such a generic treatment may expose vagueness of biological conclusions or expose unforeseen issues. A re-analysis of the relation between BLOC segments of histone modification and SINE repeats shows that conclusions regarding direct overlap at the base-pair level depends on the randomizations used in the significance analysis. Using biologically reasonable null models, the correspondence between BLOC segments and SINE repeats appears not to be due to overlap at the base-pair level, but rather seems to be due to local variation in intensities of both tracks. This does not directly oppose the original conclusions, but brings further insight into the nature of the relation. Similarly, an analysis of the relation between DNA melting and exon location confirms the conclusion from previous studies that exon boundaries coincide with gradients of melting temperature. However, taking GC content into account as a possible confounder, the analysis does not suggest a direct functional relation between melting and exons. Instead, it suggests that the association is due to the relationship of both exons and melting tracks to GC content.

We believe the generic concepts and challenges identified by our work will trigger community efforts to improve genome analysis methodology. The Genomic HyperBrowser demonstrates the feasibility of applying our approach to large-scale genomic datasets, providing a concrete basis for further research and development in inferential genomics. We thus consider the solutions presented here more like a start than an end of this important endeavor.

## Materials and methods

### Statistical methods

A track is defined over the whole genome or only in parts of it, masking away the rest. In a local analysis, statistical tests are performed in each bin with sufficient sample size. Resizing of bins allows for localization of events (similarities, differences, and so on, between the two tracks) with flexible precision. Preservation rules leads to conditional *P*-values that are not necessarily ordered, even if the preservation mechanism is incremental. Statistical tests have been tried on simulated data, also when model assumptions are not completely fulfilled. Standard Monte Carlo requires deciding on the number of Monte Carlo samples. We suggest at least two to five times the number of tests, in order to allow for FDR adjustment. Additionally, we adopt sequential Monte Carlo, where the algorithm continues sampling until the observed statistic has been exceeded a given number of times (say 20) [9]. This gives better estimates of small *P*-values with overall reduced computations. Intensity tracks are used to define non-standard null hypothesis. Several strategies for building intensity curves are described in Section 3 in Additional file 1. Intensity curves allow performing randomizations that mimic another track (or a combination of tracks), useful to account for confounding effects. For unmarked points, the intensity curve can be any regular function $\lambda_0(b)$ where b is the position along, say, a chromosome. If $\lambda_0(b) = c$ (constant), points are uniformly distributed. As another example, $\lambda_0(b)$ can be a kernel density estimate based on the track of observed points. In general, the intensity $\lambda_0(b)$ may depend on several different tracks $g_1, g_2, ..., g_k$, through a function s, so that $\lambda_0(b) = s(g_1(b), g_2(b), ..., g_k(b))$, for example, $\lambda_0(b) = c + \Sigma\beta_i g_i(b)$. An important case that requires a special choice of intensity track is when the comparison between two tracks $T_1$ and $T_2$ might be confounded by a third, confounder, track $T_3$. This is discussed in further detail in Section 5 in Additional file 1 for the melting-exon example, where each track depends on a function of the GC content.

### Software system

The Genomic HyperBrowser [30] is implemented in Python [22], version 2.7. It runs as a stand-alone application tightly connected to the Galaxy framework [2], using the version dated 2010-10-04. The user interface is based on Mako templates for Python [32], version 0.2.5, and Javascript library Jquery [33], version 1.4.2. The software uses NumPy [34], version 1.5.1rc1, for disk based vector mapping and fast vector operations. R [35], version 2.10.1, is used for plotting and basic statistical routines, using the RPy API [36], version 1.0.3. The software is open source and freely available, using GPL [37] version 3, and can be downloaded from [30]. The Genomic HyperBrowser runs on a dedicated Linux server, with large computations offloaded to the Titan cluster [38].

### Biological example: histone modifications versus gene expression

Raw histone modification data [39] were preprocessed using the NPS (Nucleosome Positioning from Sequencing) software [40], using peak detection, leading to nucleosome positioning information as short segments, treated as unmarked points (UP). Raw microarray expression values [41] were used to represent gene expression, in line with [6]. Direct comparison of the expression levels of individual probes is not generally justified. As Barski *et al.* [6] compares sets of 1,000 genes each, the direct comparison of values between groups may be justified by noise averaging (although not discussed in [6]). Using Kendall's rank correlation test, a similar reduction of error is obtained. Detailed correlation values for the different histone modifications are given in Table S1 in Additional file 1. The distribution of histone modifications relative to TSS is given for two different modifications in Figure S4 in Additional file 1.

### Biological example: histone modifications versus repeating elements

ChIP-seq data on histone modification [39,42] were preprocessed using the SICER software [43], which returns clusters of neighboring nucleosomes as islands unlikely to have appeared by chance, using an appropriate random background model. These clusters are treated as unmarked segments (US). The ChIP-chip data of H3K27me3 positions were obtained directly from Pauler *et al.* [8], and were preprocessed by them using their BLOCs software, which returns broad local enrichments, also treated as unmarked segments (US). Detailed overlap results between repeats and different histone modification sources are given in Table S2 in Additional file 1.

### Biological example: exons versus DNA melting

The melting fork probability tracks $P_L(x)$ and $P_R(x)$ used in this study were obtained using the Poland-Scheraga model [44]. To make the correction for GC content, a pair of GC-based function tracks, L(x) and R(x), were created using a moving window approach. Let $E_L$ ($E_R$) be the left (right) exon boundaries. For testing the melting-exon relation in tracks ($E_L$, $P_L$), an intensity track was created based on L(x), R(x) and $E_L$ (and similarly for tracks ($E_R$, $P_R$)). See Section 5 in Additional file 1 for more details.

## Additional material

**Additional file 1: Supplementary material**. Miscellaneous supplementary material: gene coverage example. On the importance of realistic null models. On mathematics of genomic tracks. On system architecture. On Exon DNA melting example. Supplementary figures and tables.

**Additional file 2: Statistical tests**. Detailed description of the statistical tests implemented in the software system.

**Additional file 3: Supplementary note on simulation**. Description of basic algorithms for simulating synthetic tracks, used to assess statistical tests.

### Abbreviations

BLOC: broad local enrichment; bp: base pair; F: function; FDR: false discovery rate; kb: kilo base pairs; LINE: long interspersed nuclear element; Mbp: mega base pairs; MP: marked point; MS: marked segment; SINE: short interspersed nuclear element; SNP: single-nucleotide polymorphism; TSS: transcription start site; UP: unmarked point; US: unmarked segment.

### Acknowledgements

We gratefully acknowledge ChIP-chip data provision from Florian M Pauler, and helpful comments on the manuscript from Magnus Lie Hetland, Sylvia Richardson and Håvard Rue. Gro Nilsen is acknowledged for some plotting functions, and Peter Wiedswang for administrative assistance. We thank the Scientific Computing Group at USIT for providing friendly and helpful assistance on system administration. We also thank PubGene, Inc. for kind assistance in the development of literature tracks. Additional funding was kindly provided by EMBIO, UiO and Helse Sør-Øst. This work was performed in association with 'Statistics for Innovation', a Centre for Research-Based Innovation funded by the Research Council of Norway.

### Author details

[1]Department of Informatics, University of Oslo, Blindern, 0316 Oslo, Norway. [2]Department of Tumor Biology, The Norwegian Radium Hospital, Oslo University Hospital, Montebello, 0310 Oslo, Norway. [3]Statistics For Innovation, Norwegian Computing Center, 0314 Oslo, Norway. [4]Department of Mathematics, University of Oslo, Blindern, 0316 Oslo, Norway. [5]Centre for Cancer Biomedicine, The Norwegian Radium Hospital, Oslo University Hospital, Montebello, 0310 Oslo, Norway. [6]Institute for Medical Informatics, The Norwegian Radium Hospital, Oslo University Hospital, Montebello, 0310 Oslo, Norway. [7]Department of Biostatistics, Institute of Basic Medical Sciences, University of Oslo, Blindern, 0317 Oslo, Norway.

### Authors' contributions

GKS, AF and EH conceived the approach, GKS, SG and MJ developed the software, GKS, SG, HR, TC, VN and EH developed novel track types, IKG, LH, MH, KL, EF and AF developed the statistical concepts, GKS, SG and HR tested and validated the system, and GKS, SG, HR, ET and EH developed the biological examples. All authors participated in the manuscript development, and read and approved the final manuscript.

### Competing interests

Eivind Hovig is a shareholder of PubGene, Inc. All other authors declare that they have no competing interests.

### References

1. The ENCODE (ENCyclopedia Of DNA Elements) Project. *Science* 2004, **306**:636-640.
2. Giardine B, Riemer C, Hardison RC, Burhans R, Elnitski L, Shah P, Zhang Y, Blankenberg D, Albert I, Taylor J, Miller W, Kent WJ, Nekrutenko A: Galaxy: a platform for interactive large-scale genome analysis. *Genome Res* 2005, **15**:1451-1455.
3. Pruess M, Kersey P, Apweiler R: The Integr8 project–a resource for genomic and proteomic data. *In Silico Biol* 2005, **5**:179-185.
4. Bock C, Halachev K, Buch J, Lengauer T: EpiGRAPH: user-friendly software for statistical analysis and prediction of (epi)genomic data. *Genome Biol* 2009, **10**:R14.
5. Zhu J, Sanborn JZ, Benz S, Szeto C, Hsu F, Kuhn RM, Karolchik D, Archie J, Lenburg ME, Esserman LJ, Kent WJ, Haussler D, Wang T: The UCSC Cancer Genomics Browser. *Nat Methods* 2009, **6**:239-240.
6. Barski A, Cuddapah S, Cui K, Roh TY, Schones DE, Wang Z, Wei G, Chepelev I, Zhao K: High-resolution profiling of histone methylations in the human genome. *Cell* 2007, **129**:823-837.
7. Derse D, Crise B, Li Y, Princler G, Lum N, Stewart C, McGrath CF, Hughes SH, Munroe DJ, Wu X: Human T-cell leukemia virus type 1 integration target sites in the human genome: comparison with those of other retroviruses. *J Virol* 2007, **81**:6731-6741.
8. Pauler FM, Sloane MA, Huang R, Regha K, Koerner MV, Tamir I, Sommer A, Aszodi A, Jenuwein T, Barlow DP: H3K27me3 forms BLOCs over silent genes and intergenic regions and specifies a histone banding pattern on a mouse autosomal chromosome. *Genome Res* 2009, **19**:221-233.
9. Besag J, Clifford P: Sequential Monte Carlo p-values. *Biometrika* 1991, **78**:301-304.
10. Manly BFJ: *Randomization, Bootstrap and Monte Carlo Methods in Biology* Boca Raton, FL: Chapman and Hall; 2007.
11. Jost D, Everaers R: Genome wide application of DNA melting analysis. *J Phys Condensed Matter* 2009, **21**:034108.
12. King GJ: Stability, structure and complexity of yeast chromosome III. *Nucleic Acids Res* 1993, **21**:4239-4245.
13. Liu F, Tostesen E, Sundet JK, Jenssen TK, Bock C, Jerstad GI, Thilly WG, Hovig E: The human genomic melting map. *PLoS Comput Biol* 2007, **3**:e93.
14. Suyama A, Wada A: Correlation between thermal stability maps and genetic maps of double-stranded DNAs. *J Theor Biol* 1983, **105**:133-145.
15. Yeramian E: Genes and the physics of the DNA double-helix. *Gene* 2000, **255**:139-150.
16. Tøstesen E, Sandve GK, Liu F, Hovig E: Segmentation of DNA sequences into twostate regions and melting fork regions. *J Phys Condensed Matter* 2009, **21**:034109.
17. Carlon E, Malki ML, Blossey R: Exons, introns, and DNA thermodynamics. *Phys Rev Lett* 2005, **94**:178101.
18. Hanai R, Suyama A, Wada A: Vestiges of lost introns in the thermal stability map of DNA. *FEBS Lett* 1988, **226**:247-249.
19. Cox DR, Isham V: *Point Processes* Boca Raton, FL: Chapman and Hall; 1980.
20. Grandell J: *Mixed Poisson Processes* Boca Raton, FL: Chapman and Hall; 1997.
21. Jenssen TK, Laegreid A, Komorowski J, Hovig E: A literature network of human genes for high-throughput analysis of gene expression. *Nat Genet* 2001, **28**:21-28.
22. Python Reference Manual. [http://docs.python.org/release/2.5.2/ref/ref.html].
23. Kent WJ, Sugnet CW, Furey TS, Roskin KM, Pringle TH, Zahler AM, Haussler D: The human genome browser at UCSC. *Genome Res* 2002, **12**:996-1006.
24. Beck K: *Test Driven Development* London: Addison-Wesley Profession; 2002.
25. Galaxy. [http://main.g2.bx.psu.edu/].
26. Lieberman-Aiden E, van Berkum NL, Williams L, Imakaev M, Ragoczy T, Telling A, Amit I, Lajoie BR, Sabo PJ, Dorschner MO, Sandstrom R, Bernstein B, Bender MA, Groudine M, Gnirke A, Stamatoyannopoulos J, Mirny LA, Lander ES, Dekker J: Comprehensive mapping of long-range interactions reveals folding principles of the human genome. *Science* 2009, **326**:289-293.
27. Wang Z, Zang C, Rosenfeld JA, Schones DE, Barski A, Cuddapah S, Cui K, Roh TY, Peng W, Zhang MQ, Zhao K: **Combinatorial patterns of histone acetylations and methylations in the human genome.** *Nat Genet* 2008, **40**:897-903.
28. **1000Genomes.** [http://www.1000genomes.org/].
29. Lieberman-Aiden E, van Berkum NL, Williams L, Imakaev M, Ragoczy T, Telling A, Amit I, Lajoie BR, Sabo PJ, Dorschner MO, Sandstrom R, Bernstein B, Bender MA, Groudine M, Gnirke A, Stamatoyannopoulos J, Mirny LA, Lander ES, Dekker J: **Comprehensive mapping of long-range interactions reveals folding principles of the human genome.** *Science* 2009, **326**:289-293.
30. Naumova N, Dekker J: **Integrating one-dimensional and three-dimensional maps of genomes.** *J Cell Sci* **123**:1979-1988.
31. Knoch TA, Goker M, Lohner R, Abuseiris A, Grosveld FG: **Fine-structured multi-scaling long-range correlations in completely sequenced genomes - features, origin, and classification.** *Eur Biophys J* 2009, **38**:757-779.
32. **Mako.** [http://www.makotemplates.org].
33. **JQuery.** [http://jquery.com].
34. Oliphant TE: In *Guide to NumPy.* Edited by: Spanish Fork UT. Trelgol Publishing; 2006:.
35. Team R: *R: A Language and Environment for Statistical Computing* Vienna: Austria; R Foundation for Statistical Computing; 2006.
36. **RPy a robust Python interface to the R Programming Language.** [http://rpy.sf.net].
37. **GPL.** [http://www.gnu.org/copyleft/gpl.html].
38. **Titan.** [http://www.notur.no/hardware/titan/].
39. Barski A, Zhao K: **Genomic location analysis by ChIP-Seq.** *J Cell Biochem* 2009, **107**:11-18.
40. Zhang Y, Shin H, Song JS, Lei Y, Liu XS: **Identifying positioned nucleosomes with epigenetic marks in human from ChIP-Seq.** *BMC Genomics* 2008, **9**:537.
41. Su AI, Wiltshire T, Batalov S, Lapp H, Ching KA, Block D, Zhang J, Soden R, Hayakawa M, Kreiman G, Cooke MP, Walker JR, Hogenesch JB: **A gene atlas of the mouse and human protein-encoding transcriptomes.** *Proc Natl Acad Sci USA* 2004, **101**:6062-6067.
42. Mikkelsen TS, Ku M, Jaffe DB, Issac B, Lieberman E, Giannoukos G, Alvarez P, Brockman W, Kim TK, Koche RP, Lee W, Mendenhall E, O'Donovan A, Presser A, Russ C, Xie X, Meissner A, Wernig M, Jaenisch R, Nusbaum C, Lander ES, Bernstein BE: **Genome-wide maps of chromatin state in pluripotent and lineage-committed cells.** *Nature* 2007, **448**:553-560.
43. Zang C, Schones DE, Zeng C, Cui K, Zhao K, Peng W: **A clustering approach for identification of enriched domains from histone modification ChIP-Seq data.** *Bioinformatics* 2009, **25**:1952-1958.
44. Poland D, Scheraga HA: *Theory of Helix-Coil Transitions in Biopolymers* New York: Academic Press; 1970.
45. Pruitt KD, Harrow J, Harte RA, Wallin C, Diekhans M, Maglott DR, Searle S, Farrell CM, Loveland JE, Ruef BJ, Hart E, Suner MM, Landrum MJ, Aken B, Ayling S, Baertsch R, Fernandez-Banet J, Cherry JL, Curwen V, Dicuccio M, Kellis M, Lee J, Lin MF, Schuster M, Shkeda A, Amid C, Brown G, Dukhanina O, Frankish A, Hart J, *et al*: **The consensus coding sequence (CCDS) project: Identifying a common protein-coding gene set for the human and mouse genomes.** *Genome Res* 2009, **19**:1316-1323.
46. Pruitt KD, Tatusova T, Klimke W, Maglott DR: **NCBI Reference Sequences: current status, policy and new initiatives.** *Nucleic Acids Res* 2009, **37**: D32-36.
47. Hubbard T, Barker D, Birney E, Cameron G, Chen Y, Clark L, Cox T, Cuff J, Curwen V, Down T, Durbin R, Eyras E, Gilbert J, Hammond M, Huminiecki L, Kasprzyk A, Lehvaslaiho H, Lijnzaad P, Melsopp C, Mongin E, Pettett R, Pocock M, Potter S, Rust A, Schmidt E, Searle S, Slater G, Smith J, Spooner W, Stabenau A, *et al*: **The Ensembl genome database project.** *Nucleic Acids Res* 2002, **30**:38-41.
48. Wilming LG, Gilbert JG, Howe K, Trevanion S, Hubbard T, Harrow JL: **The vertebrate genome annotation (Vega) database.** *Nucleic Acids Res* 2008, **36**:D753-760.
49. Yamasaki C, Murakami K, Fujii Y, Sato Y, Harada E, Takeda J, Taniya T, Sakate R, Kikugawa S, Shimada M, Tanino M, Koyanagi KO, Barrero RA, Gough C, Chun HW, Habara T, Hanaoka H, Hayakawa Y, Hilton PB, Kaneko Y, Kanno M, Kawahara Y, Kawamura T, Matsuya A, Nagata N, Nishikata K, Noda AO, Nurimoto S, Saichi N, Sakai H, *et al*: **The H-Invitational Database (H-InvDB), a comprehensive annotation resource for human genes and transcripts.** *Nucleic Acids Res* 2008, **36**:D793-799.